\documentclass[twocolumn,showpacs,preprintnumbers,amsmath,amssymb,prb]{revtex4-1}
\usepackage{graphicx}
\usepackage{dcolumn} 
\usepackage{bm} 
\usepackage{longtable}

\begin{document}

\title{Spin- and angle-resolved photoemission on topological materials}
\author{J. Hugo Dil}
\affiliation{
$^{1}$Institute of Physics, Ecole Polytechnique
F\'ed\'erale de Lausanne, CH-1015 Lausanne, Switzerland
\\ 
$^{2}$Photon Science Department, Paul Scherrer Institut, CH-5232 Villigen, 
Switzerland}

\date{\today}

\begin{abstract}
A historical review of spin- and angle-resolved photoemission on topological materials is presented, aimed at readers who are new to the field or who wish to obtain an overview of the activities in the field. The main focus lies on topological insulators, but also Weyl and other semimetals will be discussed. Further it will be explained why the measured spin polarisation from a spin polarised state should always add up to 100\% and how spin interference effects influence the measured spin texture.
\end{abstract}


\maketitle

\section{Introduction}

The classification of materials by the topology of their electronic structure has lead to a new way of thinking \cite{Haldane:2017,Kosterlitz:2017,Hasan:2010,Qi:2011,Armitage:2018}. It has, for example, revived the interest in the detailed band structure of materials that were previously considered to be lacking novelty, in order to search for signatures of their topology. Angle-resolved photoemission spectroscopy (ARPES) is the most direct tool to measure the electronic structure of a metallic or semiconducting material and, as exemplified by the papers in this special issue, has been used extensively to study topology. Because the topological response is often best observed in the edge states and their spin properties, the measurement of the spin texture of the electronic structure is of importance. This is best done by spin- and angle-resolved photoemission (SARPES), which is the main topic of this work.

The main experimental challenges will be shortly reviewed and it will be indicated where the development is going. After this the steep learning curve the community had to go through with respect to spin detection, and more importantly data interpretation, will be explored by the example of three dimensional topological insulators. Due to the presence of spin-orbit interaction (SOI) and selection rules in the photoemission process the measured spin polarization does not directly correspond to the initial state spin polarization and it will be shown how these can be related and what further information can be obtained. It will be explained why the measured spin polarization is always 100\% if all three spatial spin components are measured and incoherent effects are subtracted.

Besides 3D topological insulators also other topological phases typically possess spin-polarized edge states which can be probed by SARPES. A short overview of results obtained for systems ranging from topological Kondo insulators to Weyl semimetals and nodal line semimetals will be given and put in perspective. Before going to the experimental results it is useful to regard the general background of (3D) topological materials and present it in an generally accessible manner.

\section{Simplified view on topology in electronic structure}

In 1939 it was shown by William Shockley that, when starting from individual atoms and reducing their distance, the closing and reopening of an energy gap leads to the occurrence of a surface state \cite{Shockley:1939}. Whether this is a global or projected band gap does not play a role for the surface states, but it of course influences the general response of the system and the details of the band dispersion. In a static band structure picture this closing and reopening of a gap can be considered as a band inversion and every inversion contributes one surface or interface state. Thus for an even number of band inversions there is an even number of surface states and for an odd number of band inversions an odd number of surface states. 

If one wants to determine the number of surface states one therefore has to determine the number of band inversions. This is where the concept of topology comes into play. In scientific folklore the origin of the concept of topology is considered to be Euler's solution to the K\"oningsberg bridge problem; the question whether it was possible to cross every bridge of the city exactly once and return to the same place. Euler showed that the solution to the problem does not depend on the details of the streets, bridges, or rivers, but only on what we would now call topology. If any part of the city has an odd number of bridges connecting it to the other parts it is not possible to return to the same place. One can now classify cities on whether a closed path is impossible or possible, or whether they have somewhere an odd number of bridges or not, and refer to this as non-trivial ($\nu=1$) or trivial ($\nu=0$) topology. A similar nomenclature can now be used for band structures of crystals referring to an odd or even (including zero) number of band inversions around a given energy level throughout the Brillouin zone, and consequently an odd or even number of surface states.

\begin{figure}
	\centering
		\includegraphics[width=0.5\textwidth]{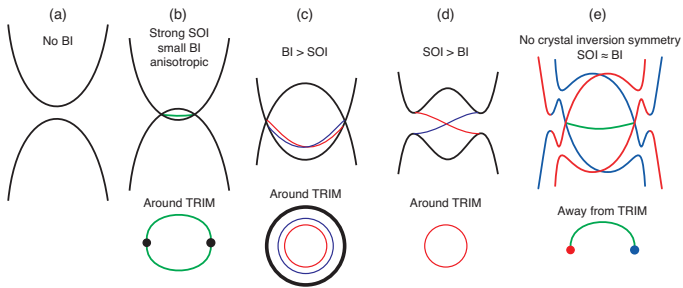}
	\caption{Schematic illustration of different topological phases based on band inversion (BI) and spin-orbit interaction (SOI) strength. From left to right the schemes illustrate a trivial semiconductor (a), a Dirac semimetal (b), a nodal line semimetal (c), a topological insulator (d), and a Weyl semimetal (e).}
	\label{topology}
\end{figure}

For a non-trivial topology the exact properties of the electronic structure around this energy level, which is typically the Fermi level, now depends on how the bulk bands are shaped after the band inversion as illustrated in Fig. \ref{topology}. The most interesting and prominent case is when the bulk bands form an absolute band gap as in Fig. \ref{topology}(d); i.e. a binding energy can be found for which no bulk band contributes to the density of states. The gap is typically due to the spin-orbit interaction (SOI) and the resultant anti-crossing of bands, but can also be due to Kondo-type or other interactions. This case with an absolute bulk band gap is referred to as a topological insulator and there will be an odd number of spin-polarised surface states crossing the gap. 

The presence of time reversal symmetry and time reversal invariant momenta (TRIM); i.e. the Brillouin zone centre and high symmetry points exactly half way between two centres, plays an important role. At these points, time reversal symmetry ($E(k,\uparrow)=E(-k,\downarrow)$) dictates that the spin polarised bands have to cross, thereby creating a Dirac cone-like dispersion of these surface states. Also mirror symmetry can force the degeneracy of spin states along certain crystal planes, resulting in cone-like dispersions away from high symmetry points in the SBZ. The latter case is relevant for topological crystalline insulators and the origin of the peculiar surface electronic structure found in them \cite{Xu:2012TCI}.

At this point it is interesting to take a slightly different look at topological insulators. Based on the concept of bulk topology described above one can consider what happens at an interface where the topology changes from non-trivial ($\nu=1$) to trivial ($\nu=0$) including vacuum. By definition the electronic structure of these insulators can not accommodate for this change by a continuous deformation, but has to go through a singularity. This singularity is a state at the interface whose existence is protected by the change in topology: i.e. it is topologically protected. In real space this topological protection can be followed in the response to defects. The interface states will move around the defects by shifting away from the interface \cite{Eremeev:2012,Pielmeier:2015,Queiroz:2016,Caputo:2016}, similar to the simplified picture for the edge state in the quantum hall effect. In momentum space, a topologically protected state has to cross the band gap and this crossing can't be lifted by small perturbations. Taking into account generic crystal symmetry this protected crossing can only be achieved by spin-polarized states because time-reversal symmetry protects their crossing as explained above. If time reversal symmetry is broken then a gap can open around the crossing point and the protection is lifted. From this it follows that the spin texture of a topological surface state is the \textit{signature} of topological protection and not the origin as often claimed.

\begin{figure}
	\centering
		\includegraphics[width=0.5\textwidth]{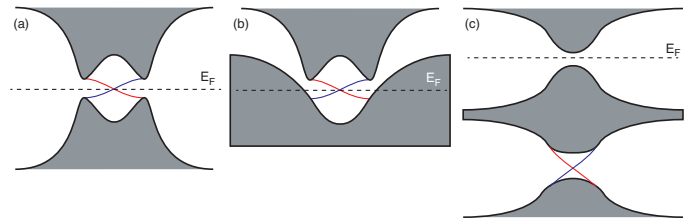}
	\caption{Influence of bulk band dispersion and gap shape on the topology. (a) Topological insulator with absolute band gap around the Fermi level. (b) Topological (semi)metal with dispersing band gap around the Fermi level. (c) System with trivial band gap around the Fermi level and an inverted (absolute) band gap in the occupied states.}
	\label{dispersion}
\end{figure}

For any application it is of course important that the topological interface states are located around the Fermi level (Fig. \ref{dispersion}(a)), but also around band gaps away from the Fermi level an odd number of band inversions can occur and thus they will be crossed by topologically protected interface states as illustrated in Fig. \ref{dispersion}(c). In any realistic material with several band gaps in the full valence and conduction band, it is very likely that several of them posses topological surface states. Of course these don't have any direct technological relevance, but studying their properties can help understand the response of materials where the topological states are located around the Fermi level.

Also if there is no absolute band gap it is possible to define the topology of the material in a similar way. The simplest case is when a gap can be traced but changes its energy strongly as a function of momentum as shown in Fig. \ref{dispersion}(b). Similar to the idea that the shape of the river will not change the solution of the Koningberg bridge problem one can straighten out the gap and compute the topology of the band structure. However, in this case it will obviously never be possible to eliminate the influence of the bulk states on the response of the system and the relevance of the topological definition will just be to determine whether there will be any surface states crossing the gap. The local band structure topology can also be calculate for a projected band gap that can't be traced throughout the Brillouin zone. If the projected band gap is due to a band inversion a pair of surface states should exist somewhere in the gap as illustrated in Fig. \ref{topology}(c). Historically these states are referred to as Shockley states as it follows his initial prediction, and famous examples are the noble metal surface states. However, in modern topological terminology the system is referred to as a nodal line semimetal and the surface states as drumhead states. At the point of writing it remained unclear whether any distinction can be made between the two different terminologies.

Another interesting case is if the bulk bands only touch in a finite collection of points. In the most general case the spin degeneracy of these points is lifted due to either broken space inversion symmetry or broken time reversal symmetry. This situation is illustrated in Fig. \ref{topology}(e) and is referred to as a Weyl semimetal in accordance with the fact that the low energy excitations in the bulk electronic structure can be described by the Weyl equation, and the crossing points of the bulk bands are referred to as Weyl points. Due to the spin texture of the bulk bands, these Weyl points come in two different flavours based on whether they are a source or drain of Berry curvature. Topologically protected surface states will connect these Weyl points of opposite character and because they start and end at different points, these surface states will take the shape of open arcs as illustrated in the figure. As will be shown below, these arcs are spin polarised and due to the low symmetry of the system they poses a complex spin texture. Perturbations to the crystal structure can reduce the band inversion and thereby cause Weyl points of opposite chirality to merge thus lifting the topological protection of the Fermi arcs.

A Dirac semimetal can be regarded as two copies of a Weyl semimetal as illustrated in Fig. \ref{topology}(b). The spin degeneracy is restored and the low energy excitations can thus be described by the Dirac equation. The open Fermi arcs now connect and form a closed contour passing through the Dirac points. It should be noted that at least two Dirac points are needed for a surface state to form. An important difference between Dirac and Weyl semimetals on the one side and nodal line semimetals and topological insulators on the other side is that in the former the bulk electronic structure is strongly anisotropic and dictated by protected crossings only along certain crystal planes and in the latter the bulk electronic structure can be regarded as isotropic in first approximation \cite{Armitage:2018}. A more detailed description of the topological phases can also be found in the other papers in this volume.

\begin{figure}
	\centering
		\includegraphics[width=0.35\textwidth]{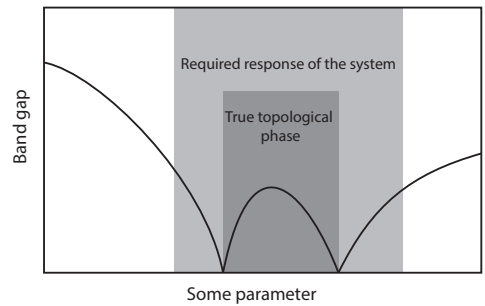}
	\caption{Illustration that the topological response, such as spin polarised surface states, can also occur outside the parameter space where the system is in a pure topological phase.}
	\label{TopResp}
\end{figure}

Topology is a mathematical concept and the separation of topological phases is thus very strict. However, the physical response of the system, such as the appearance of surface states or some transport anomaly, can vary smoothly around the transition from one topological phase to another as illustrated in Fig. \ref{TopResp}. Recently there have been several spectroscopic studies, supported by ab initio theory, illustrating this concept for a variety of topological systems \cite{Landolt:2014,Xu:2015,Tamai:2016,Russmann:2018}. Although this response might not be topologically protected, these findings indicate a larger flexibility when searching for materials with the required properties.

As a final consideration it should be realised that classification by topology is certainly useful to obtain a better understanding of the electronic structure and its response to perturbations \cite{Bradlyn:2017,Vergniory:2017}, but it should be clear that just because something can be classified by topology does not necessarily render it interesting. To make a historical comparison; the classification of elements in the periodic table has been an important scientific step forward, however, nobody will claim that an element is of interest just because it can be placed in the periodic table. For insulators the case remains as clear as indicated above, but with the ever increasing number of topological classifications for (semi)metals the question becomes what is a normal metal. Lastly, it should be realised that topological protection is only absolute in a mathematical sense. In condensed matter physics it will always be associated to an energy scale or perturbation of the crystal structure.

\section{Spin- and angle-resolved photoemission spectroscopy}

As indicated above, the presence of a single (or odd number) spin-polarised surface state is the most important signature of a topological insulator phase, and also the spin texture of the Fermi arc gives the final evidence for a Weyl semimetal phase. Therefore the possibility to measure the spin of states has played an important role in the development of the field. The most powerful experimental tool to measure the band structure of a solid is angle-resolved photoemission spectroscopy (ARPES) and the combination with spin selectivity is referred to as spin- and angle-resolved photoemission spectroscopy (SARPES).

In analogy with X-ray magnetic circular dichroism (XMCD) one could assume that measuring the band structure with opposite circular light polarisations and taking the difference will yield information on the spin texture. In it's most general sense this is called circular dichroism in angular distribution (CDAD) and such a measurement takes only about twice as long as a regular ARPES scan. However, the circular light polarisation mainly couples to the orbital angular momentum and the geometrical angular momentum and not to the spin angular momentum. Therefore CDAD primarily yields information about differences in orbital angular momentum with regard to the measurement plane and not necessarily about the spin texture. In some instances the two can coincide and CDAD can appear to give spin-resolved information \cite{Wang:2013,Bahramy:2012,Park:2012}, however in other cases states with opposite spin will look the same because they have the same orbital angular momentum \cite{Kim:2012CD,Scholz:2013}. Further studies show that the CDAD signal in topological insulators follows the point symmetry of the surface and not the spin texture \cite{Ketterl:2018}. Although CDAD can be a useful tool to draw conclusions about orbital symmetry \cite{Fedchenko:2018} or chirality \cite{Kim:2005} this shows that it should not be the method of choice to determine the spin texture of states.

A more direct SARPES approach is to actually measure the spin of the photoemitted electrons. There are several approaches to do so, based on different types of scattering effects of spin-polarised electrons from a target and comprehensive reviews can be found elsewhere \cite{Dil:2009R,Meier:2009NJP,Okuda:2013,Seddon:2016,Okuda:2017}. For the sake of this work it is sufficient to say that for every point of the band structure the expectation value of the spin polarisation of the photoemitted electron can be determined and represented as $\vec{P}=(P_x,P_y,P_z)$. Typically it is sufficient to measure the spin polarisation along a given momentum cut at a certain binding energy; a spin-resolved momentum distribution curve (MDC). Given the low efficiency of SARPES measurement full spin-resolved band maps or constant energy surfaces were not often measured, but with the advance of more efficient measurement schemes and also multiplexing of angle and energy such measurements are now becoming feasible \cite{Tusche:2013,Strocov:2015,Schonhense:2015}.

Any ARPES experiment relies on the transition from the initial to the final state, which is governed by dipole selection rules. This makes ARPES an orbital selective technique. Depending on the light polarisation and the available final state only a given number and type of orbitals can be excited, or more accurately only certain different spatial parts of the double group symmetry representation of the electronic states \cite{Tamura:1991, Irmer:1995, Irmer:1996, Yu:1998}. . Because of spin-orbit interaction a certain spin is associated with a given orbital, which thus means that the orbital selectivity has a direct impact on the measured spin polarisation. For atomic states this effect is well understood and can be used to explain a variety of effects \cite{Heinzmann:2012}. For dispersive bands the basic mechanisms are similar, but the symmetry operations and description become much more complex \cite{Henk:1994,Fanciulli:2018}. Here some of the most obvious consequences of this currently very active topic in SARPES will be covered.

In the next section an overview of the most important SARPES results obtained on topological materials will be given. However, it should be noted that the technique has provided essential information also for a wide variety of other systems that are typically not strictly classified as topological and especially where Rashba-type effects play an important role. Examples are model surface \cite{Hochstrasser:2002,Hoesch:2004,Meier:2008,Meier:2009PRB} and bulk Rashba systems \cite{Ishizaka:2011,Landolt:2012,Landolt:2015} where SARPES has given the final evidence that indeed the bands are spin split and that the spin texture follows the symmetries related to the system. In the search for systems where the spin texture can be manipulated this research has been extended to thin films \cite{Shikin:2008,Hirahara:2007s,Dil:2008,Slomski:2011B} eventually leading to the possibility to control the Rashba effect by the doping level of the semiconductor substrate \cite{Slomski:2013}. In monolayer coverages of heavy elements on semiconductors intriguing spin textures have been found \cite{Gierz:2009,Yaji:2010,Brand:2017} in some cases combined with unconventional superconductivity \cite{Matetskiy:2015} or other correlations \cite{Tegenkamp:2012,Brand:2015,Jager:2018}. Promise of applications comes from the spin textures measured in the 2DEG on isolating SrTiO$_3$(001) \cite{Santander:2014} in ferroelectric GeTe \cite{KrempaskyPRB:2016,Elmers:2016} and multiferroic (Ge,Mn)Te \cite{Krempasky:2016}. SARPES can be combined with  \textit{operando} techniques to follow the spin texture of a ferroelectric as function of applied voltage \cite{Krempasky:2018}. On spin-degenerate states SARPES can be used to extract the time scale of the photoemission process \cite{Fanciulli:2017,Fanciulli:2017B}. Furthermore, using a combination of SARPES and circular dichroism it is possible to extract the influence of spin-orbit interaction on the band structure and to examine spin-singlet and triplet type contributions in superconductors \cite{Veenstra:2014} All this shows that the applications of SARPES extend far beyond the study of topological materials and that interesting spin textures can also be found elsewhere.

\section{SARPES on topological insulators}

\begin{figure}
	\centering
		\includegraphics[width=0.45\textwidth]{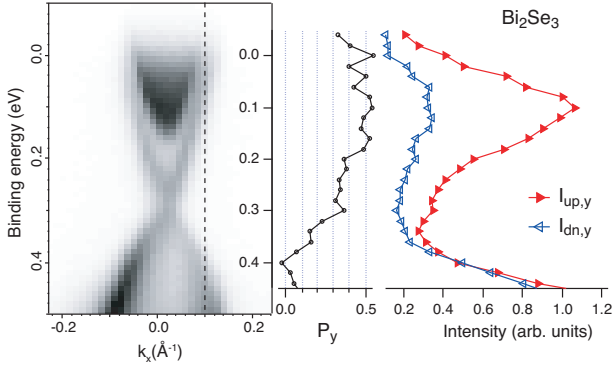}
	\caption{Band structure (left) and spin-resolved EDC (right) for the topological surface state of Bi$_2$Se$_3$ at $h\nu$=19.5eV and room temperature. Only the $P_y$ component is shown for clarity together with the back-calculated spin-resolved intensities projected on the $y$ direction. Previously unpublished data obtained in 2010 using the COPHEE end station.}
	\label{TI_spin}
\end{figure}

As explained above, one of the most outstanding properties of a 3D topological insulator is the presence of an odd number of spin-polarised surface states. The first contribution of SARPES to the study of topological insulators therefore was to measure the spin polarisation of the surface states and thereby unambiguously identify the non-trivial topology of the system. For the first generation of TIs, based on Bi$_x$Sb$_{1-x}$ alloys \cite{Hsieh:2008}, the spin polarisation could be verified \cite{Hsieh:2009}, but the large number of states (5) and the intrinsic broadening due to the alloying hindered an exact determination of all the associated properties. The next generation of topological insulators was formed by Bi$_2$Se$_3$ and related compounds and those were predicted to host a single spin-polarised surface state around the zone centre \cite{Zhang:2009}. This allowed for a clear identification of the helical spin texture by means of SARPES and thereby verify the general theory underlying topological insulators \cite{Hsieh:2009N}. In Fig. \ref{TI_spin} a representative band map and spin-resolved energy distribution curve (EDC) for Bi$_2$Se$_3$ are shown. The Dirac cone like dispersion of the topological surface state and the $P_y$ spin signal are clearly resolved.

The next open question concerned the degree of spin polarisation of the states. In a SARPES experiment the measured spin polarisation is strongly influenced by the peak to (unpolarised) background ratio, the overlap of states with different spin polarisation in the experiment, and thus by the sample quality and the experimental resolution. Furthermore, one has to take all three spatial component of the polarisation vector into account. This can be done by simultaneously fitting the three components of the spin polarisation ($P_x,P_y,P_z$) and the total intensity \cite{Meier:2008,Meier:2009NJP}. Such data analysis repeatedly showed that the degree of spin polarisation was around 100\%, both for TIs and Rashba systems, even if the measured maximum spin polarisation signal could vary.

\begin{figure}
	\centering
		\includegraphics[width=0.5\textwidth]{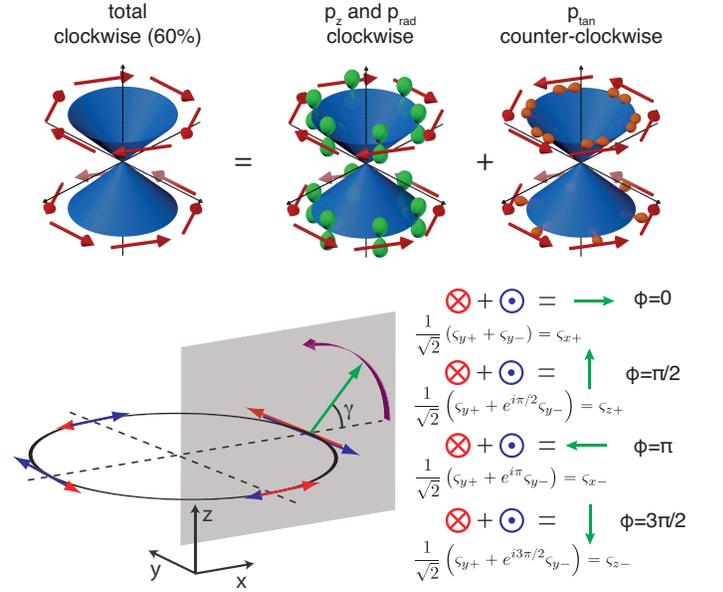}
	\caption{Schematic of the spin interference process in photoemission. The surface state of Bi$_2$Se$_3$ can be represented in terms of the $p_{rad}$, $p_{tan}$, and $p_z$ orbitals and their coupled spin textures. In the photoemission a coherent superposition of orbitals and thus spinors will be excited, resulting in a rotated spinor. The rotation angle $\gamma$ will depend on the relative phase $\phi$ of the original spinors. Figure partially adapted from \cite{Cao:2012arXiv} and \cite{Meier:2011}.}
	\label{Interference}
\end{figure}

The finding of completely spin-polarised states is unexpected at first sight because due to spin-orbit interaction spin is not a good quantum number anymore and the calculated intrinsic spin polarisation of the states is around 60\% \cite{Yazyev:2010}. For a better understanding it is helpful to rewrite the eigenfunctions in terms of the total angular momentum $J=L+S$ as done for the surface states of Bi$_2$Se$_3$ in Ref. \cite{Zhang:2013}. Using coupling parameters from ab-initio calculations one obtains that the out-of-plane $p_z$ orbitals and the radial $p_r$ orbitals are coupled to clockwise spin helicity and the tangential $p_t$ orbitals to counter-clockwise spin helicity as illustrated in Fig. \ref{Interference}. Because of the different prefactors this leads to a net clockwise spin helicity of about 60\%. The dipole selection rules in photoemission select different orbital components, including the corresponding spin component. The clearest example of this is that the measured spin helicity changes sign when using $s$- or $p$-polarised light because they probe different orbital components \cite{Jozwiak:2013,Xie:2014,Cao:2012arXiv}. 

The measured full spin polarisation can now be understood by considering that only a single orbital component is excited and thus also only a single type of spin. However, this is a strong simplification because states are hybridised and for geometrical reasons almost never only a single orbital component is excited. At this point it is important to realise that photoemission is a coherent process; if different orbitals are excited, the wave function of the photoelectron will be formed by a coherent superposition of these orbital components. Thus also the spin of the photoelectron will be a coherent superposition of the spinors related to these orbitals. In a simplified scheme using the $y$ direction as the basis, the spinors $\varsigma$ can be written as: 
\begin{align*}
\varsigma_{x+}&=\frac{1}{\sqrt{2}}\left( \begin{array}{c} 1 \\ 1 \\ \end{array} \right),\ \varsigma_{x-}=\frac{1}{\sqrt{2}}\left( \begin{array}{c} 1 \\ -1 \\ \end{array} \right) \\ 
\varsigma_{y+}&=\left( \begin{array}{c} 1 \\ 0 \\ \end{array} \right),\ \varsigma_{y-}=\left( \begin{array}{c} 0 \\ 1 \\ \end{array} \right) \\
\varsigma_{z+}&=\frac{1}{\sqrt{2}}\left( \begin{array}{c} 1 \\ i \\ \end{array} \right),\ \varsigma_{z-}=\frac{1}{\sqrt{2}}\left( \begin{array}{c} 1 \\ -i \\ \end{array} \right)
\end{align*}
Where the subscript indicates whether the spinor is parallel or antiparallel to the respective spatial direction. As illustrated in Fig. \ref{Interference} adding two spinors, or more precisely the coherent superposition of two spinors, leads to a spin state in the plane perpendicular to both, whereby the exact orientation depends on the relative phase of the spinors. In SARPES on dispersing states this spin interference was first observed in the overlap of Rashba states from the Sb/Ag(111) surface alloy \cite{Meier:2011}. Interference effects between orbitals of the same state are however easier to consider because by definition the components have the same energy and momentum and are thus coherent. 

For the topological insulator Bi$_2$Se$_3$ the partial spin reorientation along the plane perpendicular to the helical spin component was initially interpreted as an interference between contributions from different atomic layers \cite{Zhu:2013,Zhu:2014}. However, this is a inadequate picture because it assumes that the wave function is localised on different atomic sites and that photoemission is a spatially resolved process. In theory it is of course always possible to project the spin polarisation on atomic layers, as was initially done for surface states \cite{Bihlmayer:2006} and quantum well states \cite{Dil:2008} and later also for topological insulators \cite{Henk:2012,Eremeev:2012}. This can be very useful to understand the origin of spin textures and determine strategies for changing them, but it should not be confused with spatial spin modulations for a delocalised state with an extended wave function. As described above, the spin is coupled to the orbital component and this in turn is derived from the atomic contributions, but in this context space is not a property of the wave function.

\begin{figure}
	\centering
		\includegraphics[width=0.5\textwidth]{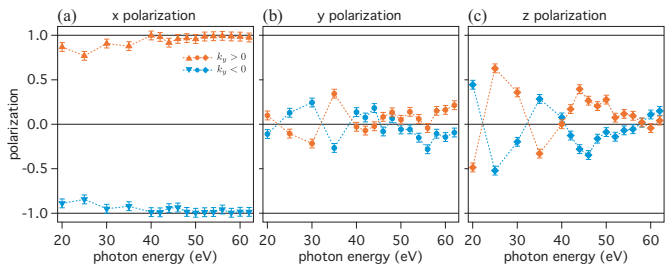}
	\caption{Dependency of the measured spin polarisation for the topological surface state of Bi$_2$Se$_3$ as a function of photon energy. The sample was tilted perpendicular to the photoemission scattering plane to access the Fermi level crossings of the state. The polarisation along the x-direction (a), y-direction (b), and z-direction (c) of the sample reference frame was determined by fitting the data to remove the background and experimental broadening.}
	\label{BiSeLandolt}
\end{figure}

The exact orbital selectivity in ARPES depends on the light polarisation, the experimental geometry, and the symmetry of the final state, but in any ARPES experiment a combination ($\ge 1$) of orbitals is probed in a coherent fashion. The final properties of some observable is formed by the coherent sum of all the contributions; i.e. including interference effects. For the spin this means that the spin texture related to every orbital component is coherently added to the other components, leading to a new spin direction depending on the relative phase. The length of this final spin vector remains 1 because each orbital contribution has a well defined spin and the photoemission process is coherent by definition \cite{Park:2012L}. It is important to note that diffraction or scattering at the surface will not change the modulus of the spin polarisation if a fully polarised beam is considered \cite{Kessler:1985}. However, the surface could induce an additional rotation of the spin polarisation vector, although up to now no experimental evidence for a significant rotation has been found \cite{Fanciulli:2018}. Therefore this effect will not be considered any further in this work.

In the interference process the phase of all the different contributions is given by the complex transition matrix element and thus depends on photon energy and experimental geometry. An example of the changing phase, for a fixed geometry and light polarisation, as a function of photon energy is given in Fig. \ref{BiSeLandolt}. Here $P_x$ is the tangential, or helical, spin component. It can be seen that there is a significant spin component in the plane perpendicular to this ($P_y, P_z$), and that the total length of the spin polarisation vector adds up to one. This perpendicular component is the result of spin interference between the different orbital components. As a function of photon energy the $P_y$ and $P_z$ component are seen to vary significantly and even change sign. These changes don't follow the proposed depth dependency, but are a result of the change of the phase of the transition matrix elements as a function of photon energy.

\begin{figure}
	\centering
		\includegraphics[width=0.4\textwidth]{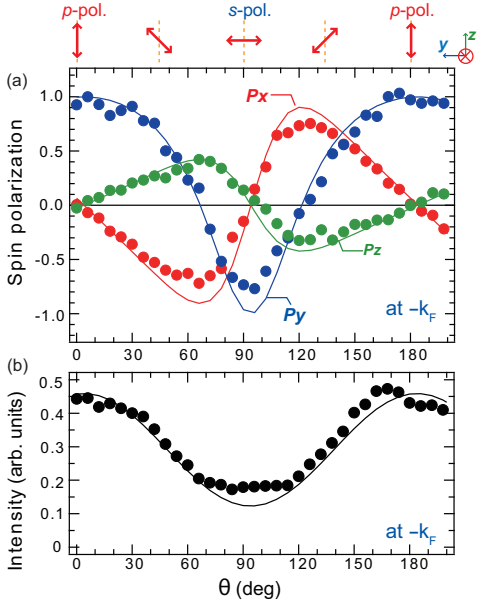}
	\caption{(a) Measured spin polarisation along the three spatial components for the surface state of Bi$_2$Se$_3$ as a function of light polarisation. (b) corresponding total intensity of the surface state. Figure adapted from \cite{Kuroda:2016}.}
	\label{Kuroda}
\end{figure}

For a fixed photon energy and experimental geometry the phase is also fixed. In this case the relative magnitudes of the different orbital contributions can be changed by varying the light polarisation. A beautiful example of this for Bi$_2$Se$_3$ can be found in Fig. \ref{Kuroda} reproduced from Ref.\cite{Kuroda:2016}. As explained above, s- and p-polarised light lead to a reversal of the measured $P_y$ spin helicity of the topological surface state of Bi$_2$Se$_3$ and when corrected for the incidence angle of the photon beam the measured polarisations are equal to 1 in these extremes. For a light polarisation rotated $45^\circ$ in between, the orbital contributions to the spin have to be added in a 1:1 ratio leading to a spin polarisation vector pointing completely in the perpendicular ($x,z$) plane, with the angle given by the phase difference between the transitions. For any other linear light polarisation this angle stays the same, but the relative magnitudes are not equal and the spin vector thus points more to $\pm P_y$ depending on whether the light polarisation angle is closer to s or p. Interestingly, in this experiment the initial state spin polarisation could be extracted from the measured difference in spin-integrated intensity between the s- and p-polarised light and was found to correspond well to theoretical predictions.

For circular polarised light there is a phase difference between the s and p components of exactly $\pm\pi/2$. Starting from from the helical spin polarisation along the $y$-direction, a coherent sum with equal magnitude and phase difference $\pm\pi/2$ leads to a spin polarisation pointing exactly along the $\pm z$-direction. Thus the measured spin polarisation will be along $P_z$ for right hand circular polarised light and along $-P_z$ for left hand circular polarised light, which can also be derived from symmetry arguments \cite{Park:2012L}. This has been confirmed by laser based SARPES experiments where such a spin reversal was observed for the surface state of Bi$_2$Se$_3$ \cite{Jozwiak:2013}. However, this simple argument is only valid if there is no additional phase difference between the transition matrix element for the s and p light component from the geometry and final state symmetry. In other words, it is only valid for normal incidence circular polarised light and under assumption of a free electron like final state. This more complex situation is also clear from the experimental results and one-step photoemission theory obtained under realistic conditions at typical photon energies above 20eV \cite{Sanchez:2014}.

These spin interference effects have to be taken into account when interpreting SARPES data on topological materials and any other system with spin-polarised initial states, and can be used to explain deviations from the predicted ground state spin texture. It is in most cases difficult to \textit{a priori} predict the phase contribution as a function of photon energy because this requires the computation of the full complex transition matrix elements. Furthermore, one has to take the projection of the photon E-field on the sample surface into account to accurately determine the magnitude of the prefactors \cite{Razzoli:2017}. Both of these considerations are implemented in one-step photoemission calculations, which are therefore capable to reproduce these effects with high accuracy.

For spin-degenerate initial states similar interference effects occur and responsible for the observation of a clear spin polarisation in SARPES \cite{Henk:1994,Heinzmann:2012}. However, here the situation is slightly more subtle: the interference between transition matrix elements is in this case responsible for the measured spin polarisation itself and not only a rotation. This goes beyond the scope of this paper and at this point it should be used as a warning that a measured spin polarisation signal in SARPES does not necessarily imply an initial state spin-polarised band. To draw this conclusion further analysis is needed and for example the observation of a 100\% spin polarisation signal in 3D SARPES is a very strong indication of spin-polarised initial states. A more detailed discussion of SARPES on spin degenerate initial states and their use can be found elsewhere \cite{Fanciulli:2018}.

The spin interference effects described above can create a spin polarisation that deviates from the expected purely helical spin texture, but also the initial state can obtain perpendicular components in the spin texture due to coupling to the crystal lattice as was already predicted and observed for Rashba systems \cite{Ast:2007,Meier:2008}. An elegant explanation for the out-of-plane and radial spin texture is given by the incorporation of higher order terms in $k$ \cite{Fu:2009w}. These higher order terms induce a deviation from a perfectly circular constant energy surface to a so-called warped, or star like, shape and at the same time a coupling between momentum and the out-of-plane spin component. For even stronger warping effects also radial terms will occur in the initial state spin texture \cite{Hopfner:2012}. In all cases this spin texture has to reflect the symmetry of the crystal structure, which imposes strong limitations. Taking into account that the spin in a pseudo vector; i.e. after a mirror operation the vector should be flipped, out-of-plane ($P_z$) spin textures are only possible for systems with odd rotational symmetry. Furthermore, this means the spin vector has to be perpendicular to a mirror plane when crossing it. Lastly, this initial state spin texture should follow the crystal symmetry and not only the measurement geometry. This means that different points in reciprocal space should be measured with the same experimental geometry to be able to distinguish photoemission induced spin effects as described above from the initial state spin texture.

\begin{figure}
	\centering
		\includegraphics[width=0.5\textwidth]{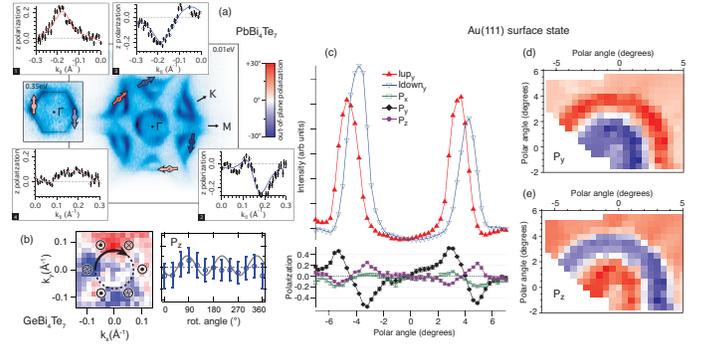}
	\caption{(a) Fermi surface and measured $P_z$ along selected azimuthal directions for the topological surface state of PbBi$_4$Te$_7$, adapted from \cite{Eremeev:2012}. (b) Measured $P_z$ for the topological surface state of GeBi$_4$Te$_7$ with a azimuthal cut to show the three-fold symmetry, adapted from \cite{Muff:2013}. (c-d) Measured spin polarisation of the surface state of Au(111) as a function of azimuthal angle (rotation around $\hat{z}$) of the sample. The top panel of (c) shows the spin split bands and the lower panel the spin polarisation along the three spatial components as a function of polar angle (rotation around $\hat{y}$). Spin polarisation along the (d) $y$- and (e) $z$-direction for positive polar angles and rotating the azimuthal angle.}
	\label{AuSS}
\end{figure}

The difference between the initial state spin texture and spin interference effects can be seen in the comparison of Bi$_2$Se$_3$ and the family based on Bi$_2$Te$_3$. In the later, the topological surface state is strongly warped and a clear $P_z$ signal is expected along the $\overline{\Gamma\mathrm{K}}$ direction \cite{Fu:2009w,Alpichshev:2010}. In order to obey time-reversal symmetry and crystal symmetry, this spin signal should reverse sign when rotating the sample by 60$^\circ$. This is exactly what is observed in SARPES both from pure Bi$_2$Te$_3$ \cite{Souma:2011,Xu:2011arXiv} and in the related PbBi$_4$Te$_7$ \cite{Eremeev:2012} and GeBi$_4$Te$_7$ \cite{Muff:2013} shown in Fig. \ref{AuSS}(a,b). Also for Bi$_2$Se$_3$ a clear $P_z$ is observed as already shown in  Fig. \ref{BiSeLandolt}, but in contrast to Bi$_2$Te$_3$ this does not depend on the azimuthal rotation of the sample although it does show an inversion with regard to normal emission. This is similar to what is observed for the Rashba-split surface state of Au(111) shown in Fig. \ref{AuSS}. The MDC in Fig. \ref{AuSS}(a) shows the Rashba-type splitting with a spin orientation along $P_y$ and also a clear $P_z$ signal that appears to obey time reversal symmetry. However, the measured $P_z$ for an azimuthal rotation of the crystal over a range of 120$^\circ$ in Fig. \ref{AuSS}(b) clearly shows that the out-of-plane spin polarisation does not follow the three-fold symmetry of the crystal and thus has to be assigned to spin interference effects.

Although topological insulators at the surface have edge states with fascinating spin properties, it should be realised that in essence most of them are narrow band gap semiconductors and behave as such with regard to doping and band bending effects. This is illustrated by the shifting of the bands in GeBi$_{4-x}$Sb$_x$Te$_7$ as a function of Sb doping \cite{Muff:2013}. In surface sensitive ARPES measurements a crossover from $n$- to $p$-type doping is found for $x=0.95$ whereas in bulk sensitive Seebeck measurements this crossover is found at $x=0.6$ for the same samples. This difference can be explained by a surface band bending of 170 meV, which for this specific system is larger as the band gap. For other systems with a larger band gap, a situation where both the bulk is insulating and at the surface only the surface state crosses the Fermi level can be found by careful doping \cite{Shikin:2014}. 

\begin{figure}
	\centering
		\includegraphics[width=0.5\textwidth]{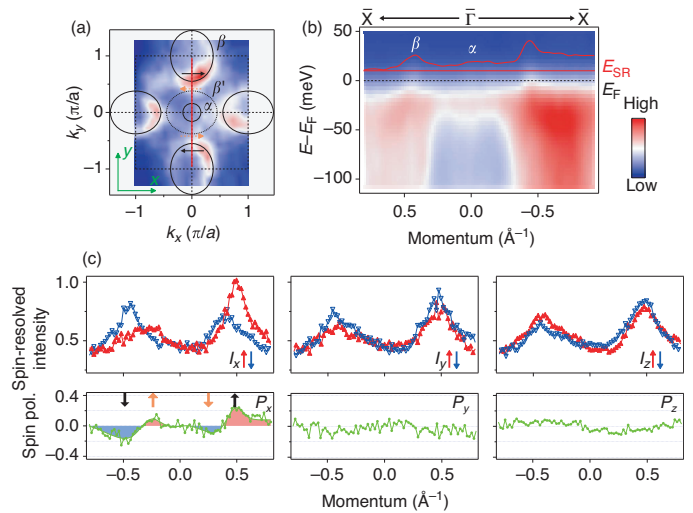}
	\caption{(a) Measured Fermi surface map of SmB$_6$(001). The ovals are guides to the eye for the surfaces states, the arrows indicate the measured spin direction. (b) Band map showing the surface states crossing the Fermi energy and the flat f-derived states. The inset shows the intensity of the surface states just above the Fermi energy. (c) Measured spin polarisation (bottom) and back-calculated intensities (top) for the three spatial components. The spin-resolved measurement in (c) was obtained along the red lines in (a) and (b). Adapted from \cite{Xu:2014}.}
	\label{SmB6}
\end{figure}

Another way to ensure that the Fermi level is in the gap throughout the sample is by letting another mechanism than spin-orbit coupling be responsible for the opening of the band gap, while still retaining parity inversion. An example of this is the topological Kondo insulator (TKI) where the Kondo effect; i.e. the hybridisation between itinerant and localised states, is responsible for the opening of a small gap \cite{Dzero:2010}. Further theoretical considerations indicated that SmB$_6$ would be a promising candidate \cite{Takimoto:2011}, which was later supported by the observation of protected surface conductivity \cite{Wolgast:2013} and ARPES measurements showing surface states on the (001) plane \cite{Neupane:2013}. Also here the final proof of the topological nature lies in the measurement of the spin texture of the surface states. Because of the small gap size and the low intensity of the surface states compared to the flat bulk states, these SARPES measurements are demanding given the lower resolution compared to regular ARPES experiments. In order to solve this problem the measurements set point for a spin-resolved MDC was set above the Fermi level and only the tail of the resolution, convoluted with the Fermi-Dirac distribution, probed the surface states \cite{Xu:2014}. The result of this measurement is reproduced in Fig. \ref{SmB6} where the spin signal along $P_x$ for the surface states becomes evident. To verify that this resembles the spin texture of the initial state, further measurements at different photon energies and polarisations, and under different geometries were performed all together establishing SmB$_6$ as a topological Kondo insulator \cite{Xu:2014}.

Recently this interpretation was called into question based on the observation of a surface state with Rashba-type spin splitting around the zone centre \cite{Hlawenka:2018}. However, this increases the number of surface state Fermi crossings by an even number and thus does not influence the topology. Furthermore, by showing how this trivial Rashba state is quenched by surface contamination whereas other surface states aren't, the paper provides another nice example of topological protection similar to what was observed in TlBiSe$_2$ \cite{Pielmeier:2015}. More recent high quality SARPES measurements on the (111) surface of SmB$_6$ show also an odd number of spin-polarised surface states in accordance to the expectation for a topological Kondo insulator that the topological surface state is present on all surfaces \cite{Ohtsubo:2018}. This appears to resolve this issue, but it does not take away the problem that SmB$_6$ is not an easy material to work with and that the involved energy scales are rather small. Therefore the search for similar materials should continue to take it beyond a proof of concept application.

In semiconductor technology a well established way to reduce the influence of band bending effects is to use thin films where the bulk to surface ratio is greatly reduced. Topological insulator materials such as Bi$_2$Se$_3$ and Bi$_2$Te$_3$ are based on quintuple layer (QL) unit cells which are stacked and bound by van der Waals forces. This makes it possible to grow high quality films of Bi$_2$Se$_3$ and related materials with QL layer precision on a variety of substrates \cite{Zhang:2010BiSe, Sakamoto:2010, Richardella:2010, Kou:2011,Tabor:2011,Schreyeck:2013}. Because the wave function of the surface state decays exponentially into the bulk over a distance of several QL, the surface states of opposite sides of the film will hybridise and open a gap for very thin films. This gap is easily observed in ARPES measurements, but the surprising aspect is that when this gap is formed the topological surface state appears to change from a Dirac-like dispersion to a Rashba-type state \cite{Zhang:2010BiSe}. 

\begin{figure}
	\centering
		\includegraphics[width=0.5\textwidth]{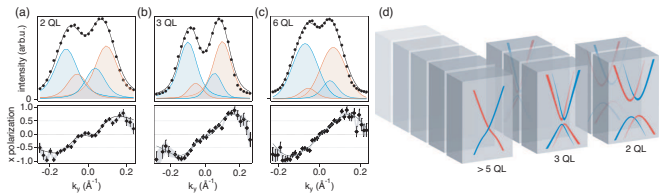}
	\caption{Measured tangential spin polarisation ($P_x$) (bottom) and fitted total intensity (top) for (a) 2 QL, (b) 3 QL, and (c) 6 QL thick films of Bi$_2$Se$_3$ grown on InP(111)B. (d) Schematic of the evolution of the band dispersion and probability density as a function of film thickness. Adapted from \cite{Landolt:2014}.}
	\label{BiSeFilm}
\end{figure}

Using a combination of SARPES and ab initio theory the exact nature of these states can be resolved \cite{Landolt:2014}. As shown in Fig. \ref{BiSeFilm} the tangential spin signal does not show large changes with film thickness when going from 2QL, where the hybridisation gap is about 300 meV, to 6QL where the gap is below the experimental resolution. The only difference is that for the thinner films a small additional wiggle is present in the spin signal around $\overline{\Gamma}$. Careful analysis of the SARPES data shows that this wiggle is due to the interface state with opposite spin helicity from the opposite side of the film. With increasing film thickness the intensity of this state diminishes and the splitting between the states at opposite surfaces becomes smaller. This is in line with theoretical considerations where one side of the film is distorted to mimic the symmetry breaking due to the substrate \cite{Landolt:2014}. By comparing the spin helicity of the outer states with what would be expected for a semi infinite film it can even be determined whether the charge transfer is from the substrate to the film or the other way around. Furthermore, the magnitude of the charge transfer is reflected in the splitting of the branches.

The Rashba-like dispersion and spin texture of the states is thus the result of the hybridisation of the interface states at opposite surfaces of the film, and in contrast to typical Rashba systems the branches are spatially separated from each other. This allows for the independent manipulation of these branches. These results also show that in thin films contributions from both surfaces will always be present in transport experiments, and will partly cancel each other with regard to spin properties. Furthermore, the SARPES results indicate that states with very similar spin properties are already present even if the system is not yet in a fully topologically protected phase.

Similar observations of the presence of spin polarised surface states on the trivial side of the transition to a topological insulator have been made using SARPES on BiTl(S$_{1-x}$Se$_x$)$_2$ as a function of sulphur doping \cite{Souma:2012T,Xu:2015}. In this case it is not the thickness but the strength of the spin-orbit interaction, or the chemical pressure, that drives the system through a topological transition around $x\approx 0.5$. On the trivial side of this transition spin polarised surface states are already present, but in contrast to the surface states in the topological phase they are gapped. Initially this gap was assigned to a ``condensed-matter version of the Higgs mechanism'' \cite{Sato:2011}, but the explanation might be less exotic based on a comparison to the Bi$_2$Se$_3$ thin films. On the trivial side the observed surface states line the projected bulk band gap and are therefore strongly coupled to the bulk states \cite{Xu:2015}. Where for the thin films the hybridisation between the surface states of opposite sides of the sample is mediated by proximity, for BiTl(S$_{1-x}$Se$_x$)$_2$ the hybridisation is mediated by the bulk states. In other words, the state with opposite spin, or the other branch of the Rashba pair, is infinitely damped by the bulk states and will be present on the opposite side of the sample. It should be noted that the presence of these states does not conflict with the concept of topological protection because they can be gapped out. On the other hand, their presence at the highly defective surface of cleaved BiTl(S$_{1-x}$Se$_x$)$_2$ indices that, in contrast to truly trivial surface states, they do experience some form of topological protection \cite{Pielmeier:2015}. Further detailed studies are required to address this issue.

These ``pre-formed'' topological surface states appear to be a general phenomena close to topological transitions. Besides the descriptions above for topological insulators they are also observed in Weyl semimetals, both by ARPES \cite{Tamai:2016,Bruno:2016,Xu:2017} and in quasi particle interference in scanning tunnelling microscopy \cite{Russmann:2018}. In this respect it should be pointed out that the observation of a single spin polarised surface state or Fermi arc is a \emph{necessary but not sufficient condition} to determine whether something is in a topological phase. For topological insulators also the Dirac point should be observed, and for Weyl semimetals also the Weyl points and how the Fermi arc connects to them. It is still an open question whether the spin properties of these ``pre-formed'' states differ from the true topological surface states and whether this will have any signature in the spin interference processes described above. Due to the orbital selectivity it is not expected that the hybridisation influences the measured degree of spin polarisation. This should still be equal to 100\% if all background influences are properly considered and if the states are separated in momentum space.

\section{SARPES on topological (semi)metals}

As explained above, also in Weyl semimetals spin polarised surface states are supposed to appear. In contrast to the surface states on topological insulators they don't form a complete contour but connect Weyl points of opposite chirality (Fig. \ref{topology}(e)). Based on this open contour they are referred to as Fermi arcs and the part to form a closed contour is located on the opposite surface connected by the bulk Weyl points. Another important difference is that the Fermi arcs don't encircle TRIM and that thus the symmetry constraints are much lower as for the surface states of topological insulators. The Weyl points of opposite chirality are found on opposite sides of a mirror plane, but for the rest their location in the Brillouin zone does not necessarily relate to any high symmetry points. This means that for the spin texture of a single Fermi arc only this mirror plane plays a role, whereas the relative spin textures of different Fermi arcs has to follow time-reversal symmetry and the crystal symmetry.

\begin{figure}
	\centering
		\includegraphics[width=0.5\textwidth]{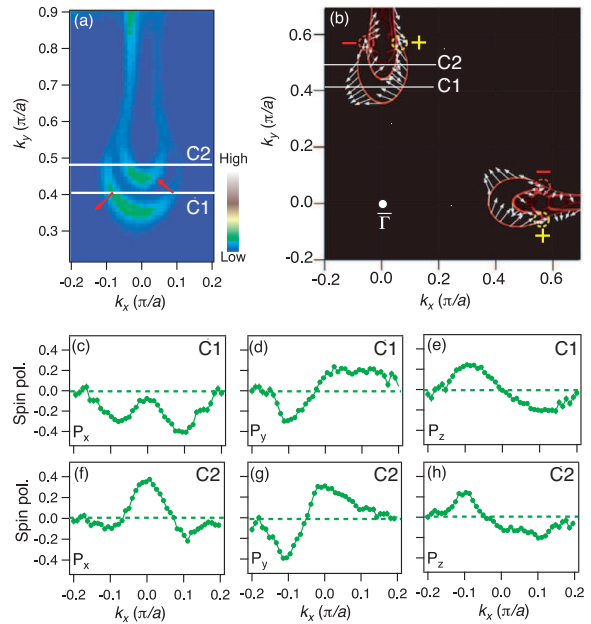}
	\caption{(a) Spin-integrated Fermi surface map for TaAs. The red arrows indicate the direction of measured in- plane spin polarizations of the Fermi arc. (b) Corresponding theoretical spin texture of surface states, with white lines indicating the locations of the SARPES measurements. Red and yellow dashed circles indicate the Weyl nodes with negative and positive chirality, respectively. (c)-(e) Measured spin polarisation along the three spatial directions along C1. (f)-(h) Same, but along C2. Adapted from \cite{Lv:2015}}
	\label{TaAs}
\end{figure}

This is nicely illustrated by the measured and calculated spin texture of the Fermi arcs of TaAs, which is one of the prototypical Weyl semimetals \cite{Lv:2015}. Fig. \ref{TaAs} shows that these spin textures are consistent with each other and that the spin is far from tangential to the constant energy contour. Only at the mirror plane the spin is perpendicular to this plane as dictated by symmetry and thus tangential to the local contour of the Fermi arc. Away from the mirror plane the spin vector rotates exactly in the opposite direction as what would be expected if it were tangential, reminiscent of the symmetry of a Dresselhaus system \cite{Zutic:2004}. The spin texture was reproduced using SARPES at low photon energies, showing the general nature of the results \cite{Xu:2016}. It should be noted that the spin texture still obeys mirror and time-reversal symmetry. By comparison to calculations this spin texture allows to identify the respective chirality of the Weyl nodes and to show that TaAs is indeed a Weyl semimetal as predicted by calculations \cite{Weng:2015,Huang:2015}.

TaAs and related compounds are so-called type I Weyl semimetals, based on the fact that the Weyl cone obeys Lorentz invariance close to the Weyl point. In type II Weyl semimetals, and in most other metals, this symmetry with regard to energy is broken and the Weyl cone becomes tilted \cite{Soluyanov:2015,Yan:2017}. As a result the Fermi surface is composed of electron and hole pockets which meet at the Weyl point in contrast to the point-like Fermi surface of a type I Weyl semimetal. Consequently, whereas the Fermi arcs of type I WSM are separated from the projected bulk band structure, the Fermi arcs from type II WSM overlap with the projected bulk band structure. Combined with the large number of bulk bands, the fact that the Weyl points are  partly in the unoccupied energy range, and the presence of ``preformed'' Fermi arcs this makes the unambigious identification of the type II Weyl semimetal phase through (S)ARPES very difficult. Even with access to the unoccupied electronic structure any possible identification remains indirect \cite{Belopolski:2016,Crepaldi:2017,Caputo:2018}. The primary candidates for type II Weyl semimetal are MoTe$_2$ and WTe$_2$ with much theoretical and experimental studies devoted to both, a review of which goes far beyond the scope of this work. Although there are several reports of conflicting conclusions, even partly by the same authors, the general consensus now appears to be that MoTe$_2$ most likely is a type II WSM whereas WTe$_2$ probably is not \cite{Bruno:2016,Crepaldi:2017,Russmann:2018}, and that for WP$_2$ the case is still open \cite{Razzoli:2018}.

These issues make SARPES on type II Weyl semimetals very challenging. On WTe$_2$ the spin texture of the (trivial) Fermi arc was measured using high resolution laser based SARPES and found to be tangential to the constant energy contour, whereas the out-of-plane component does not appear to obey the symmetry of the system \cite{Feng:2016}. For the (trivial) Fermi arc on MoTe$_2$ the measured in-plane spin polarisation is consistent with a tangential spin texture, and additionally a large out-of-plane spin polarisation is observed \cite{Weber:2018}. This out-of-plane spin polarisation indicates that the Fermi arc spin texture obeys bulk symmetry rather than surface symmetry, which would be consistent with a strongly hybridised surface resonance state \cite{Tamai:2016}. The expected non-trivial Fermi arc buried within the hole pocket has so far not been observed with SARPES.

\begin{figure}
	\centering
		\includegraphics[width=0.5\textwidth]{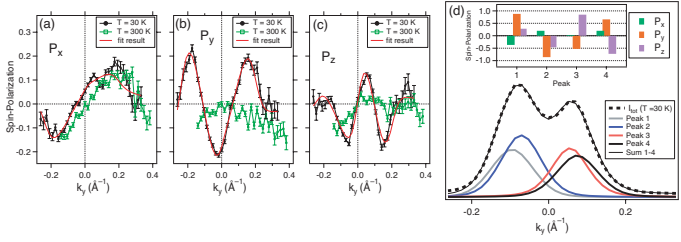}
	\caption{Measured spin-polarization along the (a) x, (b) y, and (c) z direction at the Fermi level and $k_x = 0.26$ \AA$^{-1}$ for a temperature of 300 K and 30 K. (d) Results of vectorial spin analysis for the T = 30 K data, including peak intensities and spin components (inset). Adapted from \cite{Weber:2018}.}
	\label{MoTe}
\end{figure}

In the non-magnetic Weyl semimetals discussed here the inversion symmetry is broken throughout the atomic structure of the bulk, leading to a spin polarisation of the bulk bands similar to BiTeCl \cite{Landolt:2015}. Given the large density of states of the bulk electron and hole pockets at the Fermi level it is expected that this plays a much larger role in the magnetotransport properties as the surface Fermi arcs or Weyl points. The measured spin texture of the bulk bands of both WTe$_2$ \cite{Feng:2016} and MoTe$_2$ \cite{Weber:2018} indeed show that a strongly reduced back scattering can be expected in spin conserving processes, which in turn can be suppressed by an external magnetic field. Furthermore, a careful measurement of the spin texture (Fig. \ref{MoTe}) and quasiparticle life time in MoTe$_2$ through the phase transition to a centrosymmetric system ($T_d$ to $1T^\prime$) it could be determined that new form of polar instability exists near the surface \cite{Weber:2018}.

As already indicated in the introduction, most (semi)metals can be classified by their topology and consequently spin-polarised surface states can be expected. A review of all this would constitute a review of all band structures ever measured for (semi)metals and goes far beyond the scope of this work. There are however some interesting cases that can be taken as representative for many other systems. 

The first example is tungsten. Like many transition metals, tungsten has a complex bulk band structure with a large number of projected band gaps. Depending on the number of band inversions that have created this band gap, different types of surface states can be found. Close to the Fermi level surface states with a Rashba-type spin splitting are found around the $\overline{\mathrm{S}}$ point of W(110)-H as confirmed by SARPES \cite{Hochstrasser:2002}. This is consistent with the idea that the projected band gap is only local and that the bands completely enclose the gap. On the other hand, at higher binding energy around the $\overline{\Gamma}$ point a spin polarised surface state with a Dirac cone-like dispersion can be found \cite{Miyamoto:2012}. This indicates that here the band inversion has resulted in a global gap for the involved states, even though other bulk states can be found in this gap. These results again indicate that the topological aspect of a topological insulator is not unique, but the insulating aspect is what makes a TI special.

Bismuth has long been considered as the prototypical example of a semimetal because of the low density of states at the Fermi level due to the large projected bulk band gaps. As expected a surface state can be found in this band gap of the Bi(111) surface which shows a Rashba-type spin splitting around the $\overline{\Gamma}$ point with the bands of opposite spin reconnecting with the bulk bands at the $\overline{\mathrm{M}}$ point \cite{Hofmann:2006,Hirahara:2007s}. When doping with Sb the band gap also opens at the $\overline{\mathrm{M}}$ point creating a topological insulator \cite{Hsieh:2008,Hsieh:2009}. Furthermore, ultrathin Bi(111) films of single bilayers are predicted to be 2D topological insulators with 1D edge states \cite{Liu:2011}. The extremely vicinal Bi(114) surface can be considered as a stack of Bi-bilayers looked at from the side and at each of the edges a 1D spin polarised state is formed which can be resolved by SARPES \cite{Wells:2009}. However, the bulk is not insulating and the 1D surface state partially overlaps with the projected bulk band structure, forming a 1D topological metal.

As a last example of how ubiquitous topological effects are in metals let's consider a nodal line semimetal. Similar to topological insulators, in these systems the bottom of the parabolic conduction band is shifted below the top of the valance band, but for nodal line semimetals the spin-orbit interaction is too small to open up a full gap where the bands cross. Thus the crossing of the conduction and valance band forms a closed contour and because locally the parity in inverted, a surface state can be found in this gap emerging from this closed contour. Locally the electronic structure of the bulk bands around the crossing will look like a Dirac cone along the direction perpendicular to the contour and show only little dispersion along the contour. Because of the presence of a Dirac node along a line these systems are now called nodal line semimetals, even though they have a large density of states at the Fermi level and high bulk conductivity.

The simplest example of such a system is copper, where the nodal line is located about 2 eV above the Fermi level. Also silver and gold are nodal line semimetals, whereby for gold the SOI approaches a value large enough to open a gap \cite{Yan:2015}. Because these systems follow so perfectly the example of Shockley that a gap closing and reopening results in a surface state spanning the gap, the surface states found on Cu, Ag, and Au and considered paradigm examples of Shockley surface states, which are now sometimes referred to as drumhead surfaces states. The surface state of Au(111) was one of the first to systems to be studied by SARPES \cite{Hoesch:2004} and has become the standard system to calibrate the detector or look for other effects ever since. The spin texture is tangential, although as explained above in Fig. \ref{AuSS}(b) spin interference effects play an important role. Furthermore, it was found that steps do not influence this spin texture and that also umklapp states have the same spin texture \cite{Lobo-Checa:2010}. On Cu(111) and Ag(111) the spin-orbit interaction is much smaller due to the lower Z, the different orbital contributions, and the absence of a surface reconstruction, which makes it more challenging to measure the spin texture. However, with a sensitive set-up the spin splitting of the Cu(111) surface state could be resolved in SARPES, showing a tangential spin texture and interference effects \cite{Dil:2015} the latter of which strongly depend on the defect density \cite{Fanciulli:2017}. Using a state-of-the-art laser based SARPES experiment also the spin texture of the Ag(111) surface state could be resolved with high precision \cite{Yaji:2018}.

\section{Concluding remarks}

One aim of this review was to show that surface states that are related to the topology of the bulk are more ubiquitous as they appear on first sight. Any band inversion, even if it is only locally in momentum space, will result in a surface state due to the local difference in topology compared to vacuum. All these surface states have a well defined spin texture which can, in principle, be measured by SARPES. Whether these spin-polarised surface states can be expected to significantly contribute to transport depends on the density of bulk states at the Fermi level and especially on whether a full gap is opened up. The elegance of using topology to describe band structures, and how several of the observed quasiparticles are the condensed matter counterpart of the mathematical description in high energy physics are aspects that have not received much attention in this work because the focus primarily lies on the measured spin properties. By the example of the surface states of topological insulators it was explained why the measured spin polarisation should always be 100\% if all three spin components are measured and the background and spectral overlap are considered. The basic idea of spin interference and its influence on the measured spin texture were explained and it was shown that such effects should be taken into account in all SARPES measurements. Throughout this work the sample temperature has been ignored, except for the case where it drives the system through a phase transition. The reason for this is that temperature has no influence on the measured spin texture and the other effects described here. The primary reason to use low temperatures is to enhance the spectral resolution and to reduce the incoherent overlap of states.

Lastly it should be stressed that all photoelectrons are highly spin polarised, whereby symmetry arguments can be used to discriminate different effects. The observation of a spin signal should not be used to directly draw the conclusion that the initial state is spin polarised. Similarly, the observation of a (spin-polarised) surface state is not direct evidence of the topological phase of the bulk; the so-called bulk-boundary correspondence only works in the direction that if the bulk is (locally) non-trivial a surface state must exist. In most cases the comparison to theory can resolve the question and when used with care SARPES is a powerful tool for the study of topological materials.

\section{Acknowledgements}

I am greatly indebted to C. Beenakker, E. Mele, and S. Murakami for the discussions about how to create a simplified view on topology, surface states, and band inversions without losing physical background. I take full responsibility for any over-simplifications. The SARPES results presented here and their interpretation are the result of over ten years of fruitful collaboration on this topic with a variety of external groups and of course my collaborators at the EPFL, Swiss Light Source, and University of Zurich. This work was supported by the Swiss National Science Foundation project no. PP00P2\_170591.

\footnotesize
\bibliography{references}
\bibliographystyle{apsrev4-1}


\end{document}